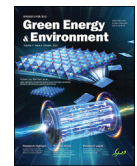

Research paper

# Green fabrication of nickel-iron layered double hydroxides nanosheets efficient for the enhanced capacitive performance

Yuchen Wang [a], Zuo Chen [a], Man Zhang [a], Yaoyu Liu [a], Huixia Luo [c,*], Kai Yan [a,b,*]

[a] School of Environmental Science and Engineering, Sun Yat-sen University, 135 Xingang Xi Road, Guangzhou, 510275, China
[b] Guangdong Provincial Key Laboratory of Environmental Pollution Control and Remediation Technology, Guangzhou, 510275, China
[c] Key Lab of Polymer Composite & Functional Materials, School of Materials Science and Engineering, Sun Yat-sen University, Guangzhou, 510275, China

## Abstract

Rational synthesis of robust layered double hydroxides (LDHs) nanosheets for high-energy supercapacitors is full of challenges. Herein, we reported an ultrasonication-assisted strategy to eco-friendly fabricate NiFe-LDHs nanosheets for the enhanced capacitive behavior. The experimental results combined with different advanced characterization tools document that the utilization of ultrasonication has a profound effect on the morphology and thickness of the as-obtained NiFe-LDHs, alternatively affecting the capacitive behavior. It shows that NiFe-LDHs nanosheets prepared with 2-h ultrasonic treatments display the exceptional capacitive performance because of the synergetic effect of ultrathin thickness, large specific surface area, and high mesoporous volume. The maximum specific capacitance of $Ni_3Fe_1$-LDHs nanosheets with the thickness of 7.39 nm and the specific surface area of 77.16 $m^2\ g^{-1}$ reached 1923 $F\ g^{-1}$, which is competitive with most previously reported values. In addition, the maximum specific energy of the assembled NiFe-LDHs//AC asymmetric supercapacitor achieved 49.13 $Wh\ kg^{-1}$ at 400 $W\ kg^{-1}$. This work provides a green technology to fabricate LDHs nanosheets, and offers deep insights for understanding the relationship between the morphology/structure and capacitive behavior of LDHs nanosheets, which is helpful for achieving high-performance LDHs-based electrode materials.

## 1. Introduction

Supercapacitors (SCs) are extreme high-power, great energy efficient and long-term cycling energy storage devices, which are widely used in portable electronic devices, electrical vehicles, and military applications [1–3]. Carbon materials are most common electrode materials for SCs. In recent years, the electrochemical performance of SCs is greatly improved with the advancement of novel carbon materials, such as hierarchical porous carbon, carbon nanotubes, graphene [4–9]. However, carbon-based SCs are still suffered from their lower specific energy (< 30 $Wh\ kg^{-1}$) [10,11]. Hence, it is urgent to develop and design high-performance electrode materials for SCs with both high specific power and high specific energy.

Layered double hydroxides (LDHs) are a kind of hydrotalcite-like two-dimensional (2D) materials, which have drawn tremendous attention as electrode materials for SCs [12,13]. The unique lamellar structure and adjustable transition metal-cation composition of LDHs give rise to extremely high theoretical specific capacitance (> 3000 $F\ g^{-1}$) [14,15]. However, LDHs usually are presented as bulk form with inevitable drawbacks, including poor electrical conductivity and insufficient exposure of active sites [16,17].

* Corresponding authors.
*E-mail addresses:* luohx7@mail.sysu.edu.cn (H. Luo), yank9@mail.sysu.edu.cn (K. Yan).

Several strategies have been utilized to overcome the disadvantages of LDHs, including exfoliation/delamination of bulk LDHs into LDHs nanosheets, combination of LDHs with highly conductive substrates, and preparation of LDHs derivatives (metallic oxides, sulfides, etc.) [18,19]. Among these strategies, exfoliation of bulk LDHs could facilitate pseudocapacitive processes by enhancing the exposed surface area and increase the electrical conductivity by removing the anions between host layers [20], which greatly improve the capacitive performance or catalytic performance of LDHs. Our group has developed Ar-etched or $NH_4F$-etched exfoliated NiCo-LDHs, CoMn-LDHs for oxygen evolution reaction [21,22], while few have investigated the capacitive performance. Recently, researchers have confirmed that the exfoliated LDHs nanosheets displayed greater electrochemical performance than the stacked LDHs. For example, Zhao et al. [23] fabricated NiTi-LDHs monolayer nanosheets through a reverse microemulsion method with the assistance of sodium dodecyl sulfate (SDS) and 1-butanol, the NiTi-LDHs nanosheets delivered an ultrahigh specific capacitance of 2310 F $g^{-1}$ at 1.5 A $g^{-1}$, whereas the bulk NiTi-LDHs possessed low specific capacitance of 393 F $g^{-1}$. Yang et al. [24] synthesized ultra-thin NiMn-LDHs nanosheets using a hydrothermal method with the aid of ethylenediamine (EN). The specific capacitance of resulting NiMn-LDHs nanosheets reached 1513 F $g^{-1}$ at 1 A $g^{-1}$, which was almost 1.7 times higher than that of bulky NiMn-LDHs. Nevertheless, these exfoliation methodologies often involve time-consuming, multi-step procedures and toxic organic solvents [25,26]. Therefore, it is significant to develop a facile and eco-friendly method to derive LDHs nanosheets with rational structure for high-energy SCs.

Ultrasonication is a simple and reliable treatment to exfoliate LDHs nanosheets by producing shock waves and microjet flows on LDHs [27]. Our previous work successfully synthesized ultrathin NiCr-LDHs nanosheets (4−5 nm) in water through one-step method with the combination of ultrasonication and mechanical stirring [28]. Compared with bulk NiCr-LDHs (without ultrasonication), the specific capacitance of ultrasonication-assisted exfoliated NiCr-LDHs nanosheets exhibited the 5-fold enhancement. However, the effects of ultrasonication on the morphology/structure and capacitive behavior of LDHs nanosheets are rarely discussed.

The present work systematically investigates the relationship between the morphology/structure and capacitive behavior of NiFe-LDHs nanosheets. NiFe-LDHs nanosheets with different morphologies were achieved by controlling the ultrasonic time. The morphology and the structure of NiFe-LDHs were characterized by a series of advanced measurements including X-ray diffraction (XRD), thermogravimetric analysis (TGA), transmission electron microscopy (TEM), atomic force microscopy (AFM), and Brunauer−Emmett−Teller (BET). The capacitive performance of NiFe-LDHs were examined as electrode materials in SCs using cyclic voltammetry (CV), galvanostatic charge/discharge (GCD), electrochemical impedance spectroscopy (EIS), and cycling tests. This work is expected to provide a clear direction for designing and optimizing the structure of the ultrasonication-assisted exfoliated LDHs nanosheets for energy storage.

## 2. Experimental section

### 2.1. Synthesis of NiFe-LDHs

All chemicals were used without any treatment and purification. NiFe-LDHs nanosheets were synthesized according to the previous reported processes [28] (Fig. 1). Nickel chloride hexahydrate ($NiCl_2 \cdot 6H_2O$, AR, 99%, Macklin) and iron chloride hexahydrate ($FeCl_3 \cdot 6H_2O$, ACS, Aladdin) powders were dissolved in ultrapure water to form 1 mol $L^{-1}$ aqueous solutions. These aqueous solutions were added to a beaker containing 100 mL ultrapure water to achieve solution A. In this work, the element molar ratio of nickel and iron was kept at 3:1 for studying the effect of ultrasonication. Solution A was treated with ultrasonication (40 kHz) and mechanical stirring (300 r $min^{-1}$) simultaneously. Afterwards, solution B, the mixed aqueous solution of 0.8 mol $L^{-1}$ sodium hydroxide (NaOH, GR, 97%, Macklin) and 0.2 mol $L^{-1}$ sodium carbonate ($Na_2CO_3$, GR, ≥ 99.9%, Macklin), was added dropwise slowly to solution A. The pH value was adjusted to 8.5 and held for 30 min. Specific time of ultrasonic treatment (0 h, 1 h, 2 h, and 3 h) was controlled and the corresponding samples were named as $Ni_3Fe_1$-LDHs-0, $Ni_3Fe_1$-LDHs-1, $Ni_3Fe_1$-LDHs-2, and $Ni_3Fe_1$-LDHs-3, respectively. The final products were collected after centrifugation and vacuum drying at 60 °C for 12 h.

### 2.2. Characterizations of NiFe-LDHs

Material and electrochemical characterizations of NiFe-LDHs are described in Supplementary Information.

## 3. Results and discussion

### 3.1. Structural and morphological characterization

The crystal structures of $Ni_3Fe_1$-LDHs were investigated by X-ray diffraction method. As shown in Fig. 2a, the characteristic diffraction peaks of the samples at 11.3°, 22.7°, 34.4°, 38.9°, and 60.1° were indexed to the (003) (006), (012) (015), and (110) crystal planes of NiFe-LDHs (PDF#40–0215), respectively, showing that these samples have typical hydrotalcite-like crystal structure [29]. It is noteworthy that the XRD patterns of the samples with less ultrasonic time (0 h, 1 h) displayed sharper and narrower peaks, implying higher crystallinity [14,30]. In other words, the utilization of the ultrasonic treatment in the preparation process demonstrates the successful exfoliation of the NiFe-LDHs into ultrathin nanosheets, which is beneficial to increase the specific surface area, enhance the exposure of redox active sites, and thus improve the capacitive performance as the electrode materials. Furthermore, the lattice parameters (*a* and *c*) of these NiFe-LDHs samples were calculated from XRD patterns (Table S1). As seen from Fig. 2b, with the increasing ultrasonic time, the *a* values remained almost the same and the $c$ values increased slightly, pointing out that the ultrasonic treatment could enlarge the interlayer spacing during the synthesis process. The increase of the interlayer spacing of LDHs promotes the diffusion of electrolyte ions, leading to enhanced electrochemical performance. TGA measurements was also performed to analyze the compositions of NiFe-LDHs. From Fig. S1, the weight losses of NiFe-LDHs below 120 °C and around 300 °C were attributed to the dehydration (absorbed water and interlayer water) and decomposition of interlayer carbonate, respectively [31].

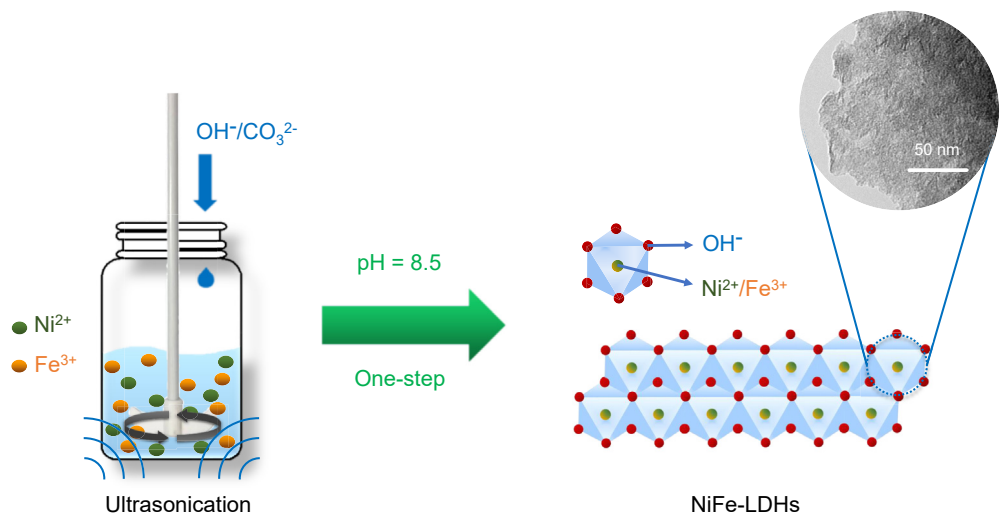

Fig. 1. One-step synthesis process of NiFe-LDHs nanosheets with the combination of ultrasonication and mechanical stirring.

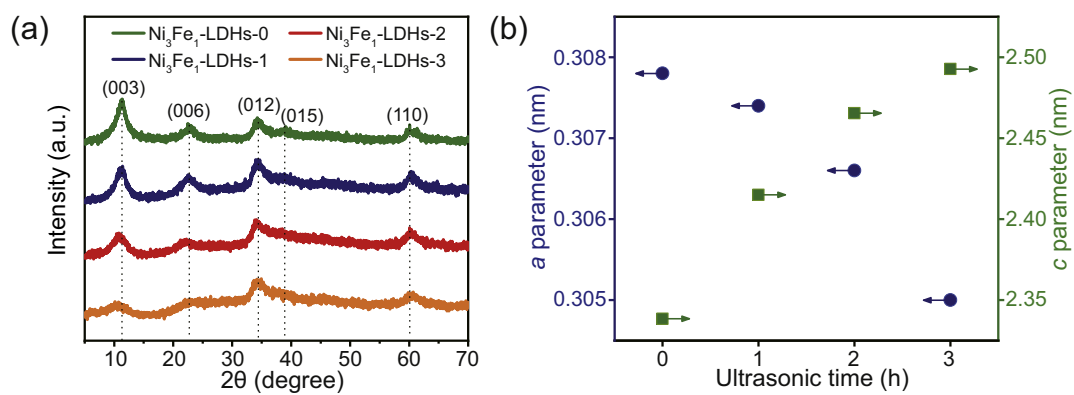

Fig. 2. (a) XRD patterns of $Ni_3Fe_1$-LDHs-0, $Ni_3Fe_1$-LDHs-1, $Ni_3Fe_1$-LDHs-2, and $Ni_3Fe_1$-LDHs-3. (b) The dependence of $a$ and $c$ parameters with ultrasonic time.

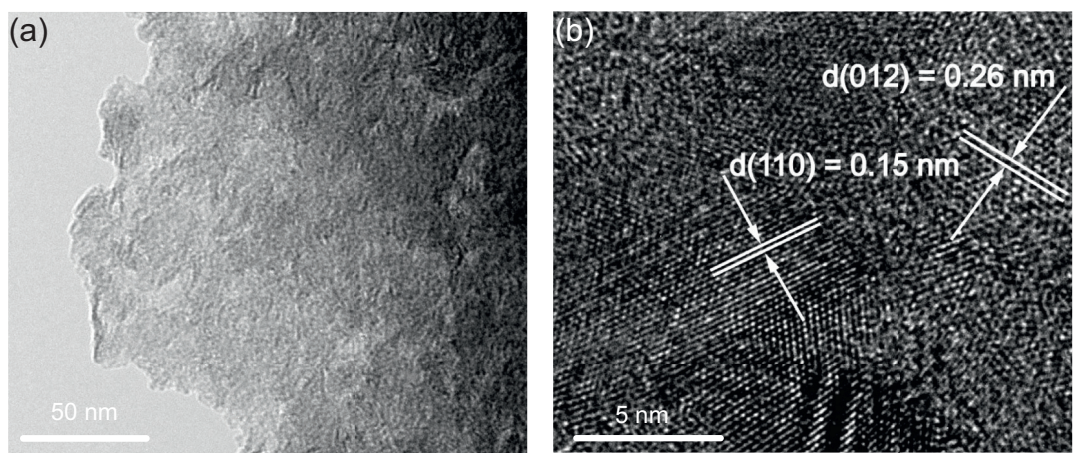

Fig. 3. (a) TEM and (b) HRTEM images of the as-synthesized $Ni_3Fe_1$-LDHs-2.

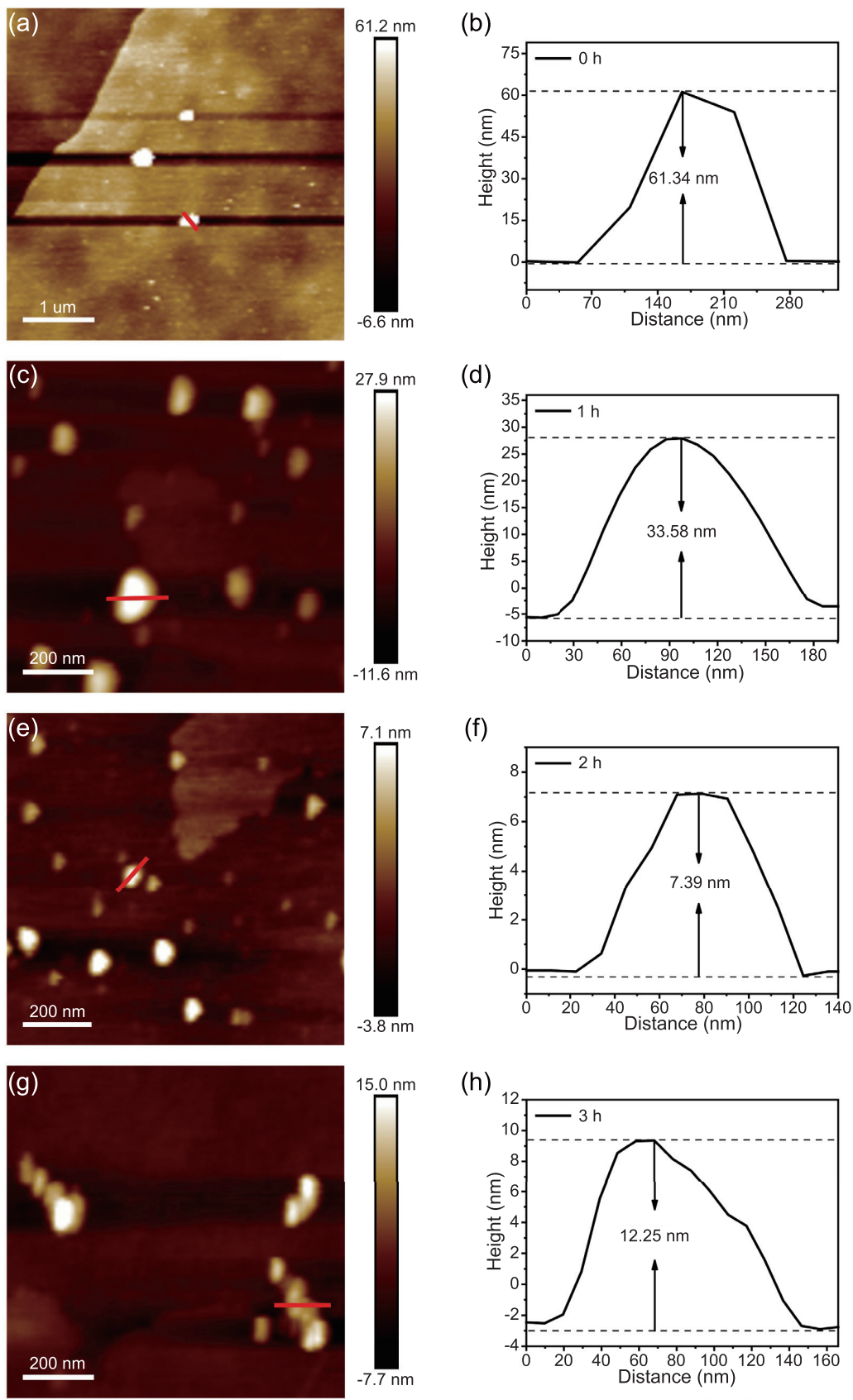

Fig. 4. AFM images and corresponding height profiles of (a, b) Ni$_3$Fe$_1$-LDHs-0 (c,d) Ni$_3$Fe$_1$-LDHs-1, (e, f) Ni$_3$Fe$_1$-LDHs-2, and (g, h) Ni$_3$Fe$_1$-LDHs-3.

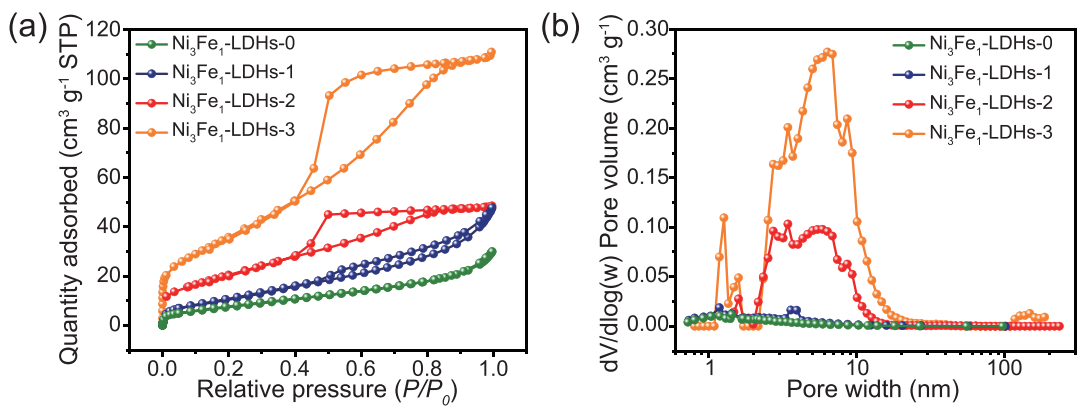

Fig. 5. (a) Nitrogen adsorption/desorption isotherms, (b) pore size distribution of Ni$_3$Fe$_1$-LDHs-0, Ni$_3$Fe$_1$-LDHs-1, Ni$_3$Fe$_1$-LDHs-2, and Ni$_3$Fe$_1$-LDHs-3.

TEM was conducted to observe the microstructure of as-synthesized NiFe-LDHs using different ultrasonic time. From Fig. S2a, the non-ultrasonic NiFe-LDHs material showed obvious bulk LDHs structure. With the increase of ultrasonic time from 0 h to 1 h, the material was partially exfoliated (Fig. S2b). When the ultrasonic time was increased to 2 h, the transparent edge of NiFe-LDHs material possessed an ultra-thin character (Fig. 3a). However, for the NiFe-LDHs material after 3 h of ultrasonic treatment, the ultrathin nanosheets were severely damaged and a large number of small fragments were generated (Fig. S2c). Moreover, the HRTEM image (Fig. 3b) presented the measured lattice fringes of 0.26 nm and 0.15 nm, which corresponded to crystal planes (012) and (110) of NiFe-LDHs in XRD results, respectively. The selected area electron diffraction (SAED) pattern of Ni$_3$Fe$_1$-LDHs-2 in Fig. S2d further confirmed the crystal structure of LDHs. The rings could be indexed to (012) (015), and (110) diffraction planes of NiFe-LDHs.

The nanosheets of Ni$_3$Fe$_1$-LDHs-2 were also analyzed using scanning transmission electron microscopy (STEM) coupled with energy dispersive X-ray spectroscopy (EDX) to determine the composition of LDHs. The EDX mappings (Fig. S3) revealed the homogeneous distribution of Ni, Fe, and O throughout the surface of Ni$_3$Fe$_1$-LDHs-2. According to Table S2, the average atomic ratio of Ni to Fe was calculated as 3.4:1, which almost matches the theoretical value.

AFM was used to measure the thickness values of NiFe-LDHs (Fig. 4 and Fig. S4). The tendency of the thickness of NiFe-LDHs with respect to ultrasonic time is consistent with TEM analyses. During the synthesis process, ultrasonic treatment greatly reduces the thickness of NiFe-LDHs. After 2 h of ultrasonication, the thickness of NiFe-LDHs decreased from 61.34 nm to only 7.38 nm. However, when the ultrasonic time was further increased to 3 h, the thickness of NiFe-LDHs was increased slight owing to negative effect of excessive ultrasonication. The forces induced by excessive ultrasonic treatment are continuously applied on the surface of NiFe-LDHs nanosheets, giving rise to severe damage on the structure of LDHs.

BET measurements was carried out to decide the specific surface area, total pore volume, and pore size, which are key factors for determining capacitive performance of NiFe-LDHs. Fig. 5a illustrates nitrogen adsorption/desorption isotherms of NiFe-LDHs. The curves with the obvious hysteresis loop exhibit type IV, indicating mesoporous dominated structures [32]. The pore distributions in Fig. 5b present that NiFe-LDHs are composed of a great quantity of mesopores. According to BET model and the Barrett−Joyner−Halenda (BJH) theory, the specific surface area, total pore volume, and average pore size are summarized in Table 1. The specific surface area and total pore volume of NiFe-LDHs nanosheets are up to 136.40 m$^2$ g$^{-1}$ and 0.178 cm$^3$ g$^{-1}$, with a significant improvement in comparison to 27.94 m$^2$ g$^{-1}$ and 0.044 cm$^3$ g$^{-1}$ for bulky NiFe-LDHs, respectively. The average pore size values further confirm the mesoporous structure of NiFe-LDHs. It is demonstrated that the ultrasonic treatment could enhance the number of active sites and facilitate the ion transport during charge/discharge processes.

### 3.2. Electrochemical performance

The electrochemical performance of NiFe-LDHs was evaluated through CV, GCD and EIS in three-electrode system. Fig. 6a shows CV curves of Ni$_3$Fe$_1$-LDHs-0, Ni$_3$Fe$_1$-LDHs-1, Ni$_3$Fe$_1$-LDHs-2, and Ni$_3$Fe$_1$-LDHs-3 electrode at 10 mV s$^{-1}$ in the potential range from −0.2 V to 0.5 V (vs. SCE). When the potential was higher than 0.5 V, the side reaction related to oxygen evolution was pronounced. The remarkable characteristic peaks observed in CV curves are mainly attributed to the redox reactions between Ni(OH)$_2$ and NiOOH [33]. Compared with other electrode materials, Ni$_3$Fe$_1$-LDHs-2 electrode possessed the largest current intensity owing to the well-defined structure of LDHs nanosheets, which was described in the previous section. The CV curves of Ni$_3$Fe$_1$-LDHs-2 electrode at different scan rates of 10–100 mV s$^{-1}$ are displayed in Fig. 6b. With the increasing scan rate, the specific current increased and thus the IR drop related to ohmic resistance and polarization resistance elevated, resulting in the shifts of the redox peaks [34].

Table 1
Comparison of BET specific surface area, total pore volume, and average pore size of Ni$_3$Fe$_1$-LDHs-0, Ni$_3$Fe$_1$-LDHs-1, Ni$_3$Fe$_1$-LDHs-2, and Ni$_3$Fe$_1$-LDHs-3.

| Sample | $S_{BET}$ (m$^2$ g$^{-1}$) | $V_{total}$ (cm$^3$ g$^{-1}$) | $D_{avg}$ (nm) |
| --- | --- | --- | --- |
| Ni$_3$Fe$_1$-LDHs-0 | 27.94 | 0.044 | 5.80 |
| Ni$_3$Fe$_1$-LDHs-1 | 40.15 | 0.071 | 5.63 |
| Ni$_3$Fe$_1$-LDHs-2 | 77.16 | 0.080 | 3.62 |
| Ni$_3$Fe$_1$-LDHs-3 | 136.40 | 0.178 | 3.78 |

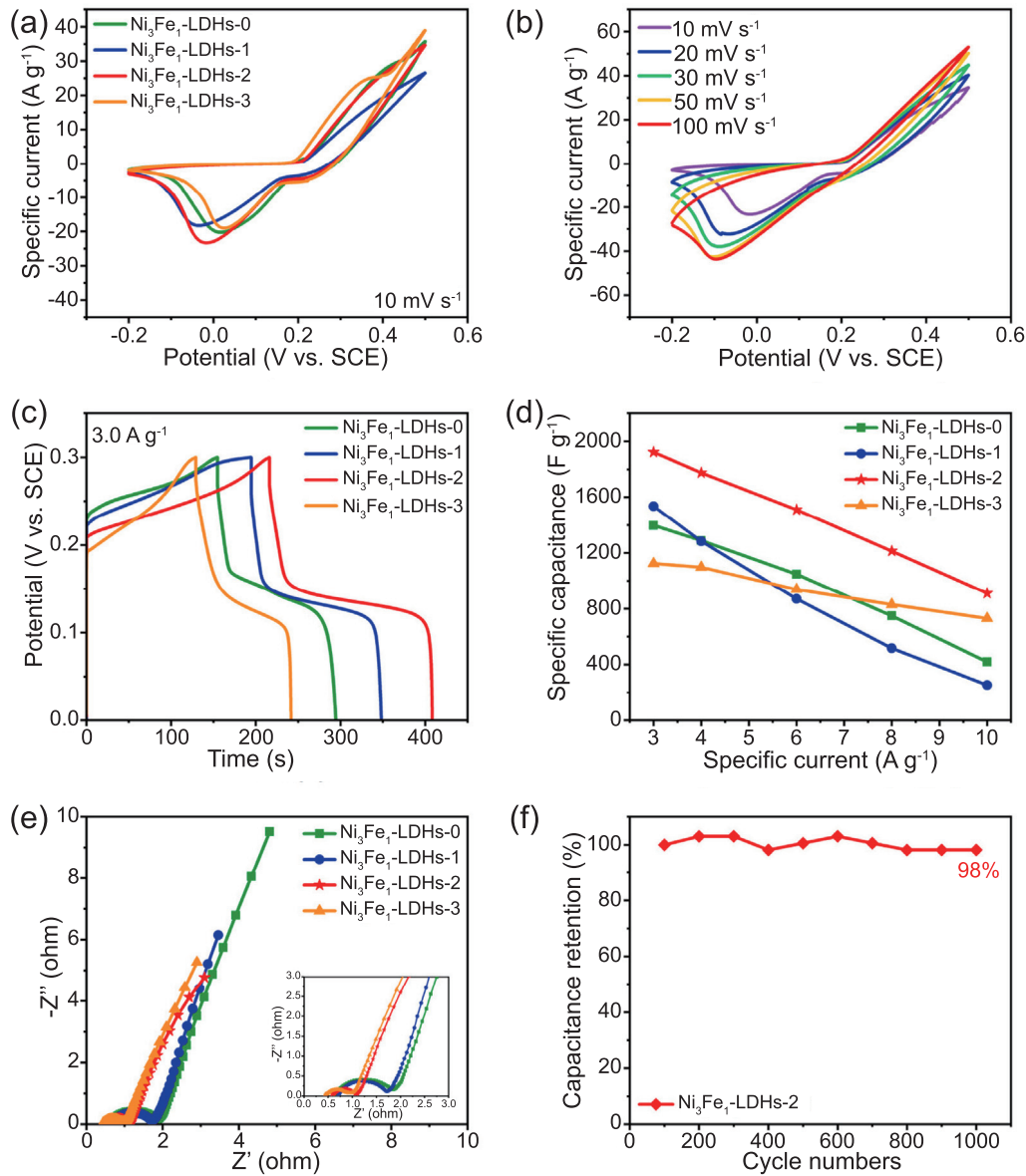

Fig. 6. (a) CV curves of Ni$_3$Fe$_1$-LDHs-0, Ni$_3$Fe$_1$-LDHs-1, Ni$_3$Fe$_1$-LDHs-2, and Ni$_3$Fe$_1$-LDHs-3 electrode at scan rate of 10 mV s$^{-1}$. (b) CV curves of Ni$_3$Fe$_1$-LDHs-2 electrode at different scan rates. (c) GCD curves of Ni$_3$Fe$_1$-LDHs-0, Ni$_3$Fe$_1$-LDHs-1, Ni$_3$Fe$_1$-LDHs-2, and Ni$_3$Fe$_1$-LDHs-3 electrode at a specific current of 3 A g$^{-1}$. (d) Specific capacitance of the Ni$_3$Fe$_1$-LDHs electrodes at different specific currents. (e) Nyquist plots of Ni$_3$Fe$_1$-LDHs-0, Ni$_3$Fe$_1$-LDHs-1, Ni$_3$Fe$_1$-LDHs-2, and Ni$_3$Fe$_1$-LDHs-3 electrode from 0.01 Hz to 100 kHz. (f) Cycle stability of Ni$_3$Fe$_1$-LDHs-2 electrode (50% conductive carbon) at 10 A g$^{-1}$.

The comparison of GCD curves of Ni$_3$Fe$_1$-LDHs-0, Ni$_3$Fe$_1$-LDHs-1, Ni$_3$Fe$_1$-LDHs-2, and Ni$_3$Fe$_1$-LDHs-3 electrode (Fig. 6c) is in good agreement with CV results. Ni$_3$Fe$_1$-LDHs-2 electrode displayed the longest discharge time. Specific capacitance values were calculated from GCD curves. Fig. S5 gives GCD curves of Ni$_3$Fe$_1$-LDHs-2 electrode at various specific currents. The symmetrical GCD curves revealed great columbic efficiency and high electrochemical reversibility. As seen in Fig. 6d, the highest specific capacitance of Ni$_3$Fe$_1$-LDHs-2 electrode reached 1923 F g$^{-1}$, which is almost 1.4 times of that of bulky Ni$_3$Fe$_1$-LDHs-0 electrode (1398.5 F g$^{-1}$). The successful exfoliation of bulk LDHs by ultrasonication is imperative to the enhancement of the specific capacitance value. However, the specific capacitance values of the electrode dropped significantly when the specific current increased to 10 A g$^{-1}$. The obvious degradation of specific capacitance is mainly caused by insufficient hydroxide ions for redox reactions and great IR drop at high specific currents [35].

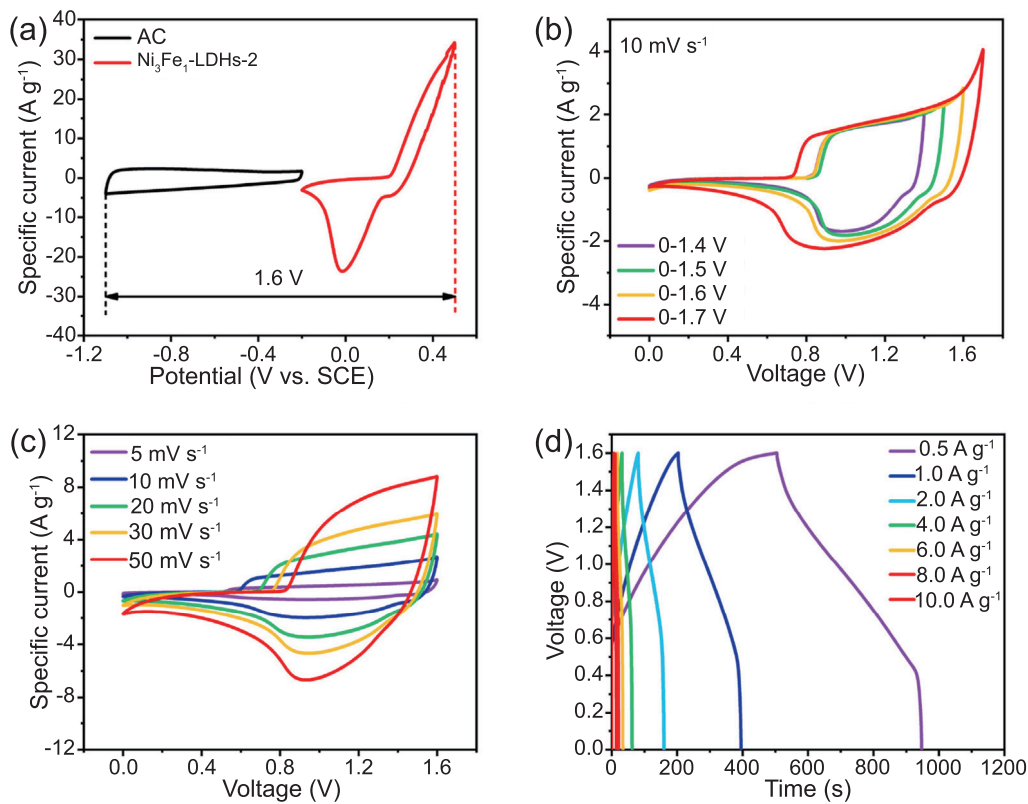

Fig. 7. (a) CV curves of Ni$_3$Fe$_1$-LDHs-2 and AC at the scan rate of 10 mV s$^{-1}$, (b) CV curves of ASC performed at different voltage windows at 10 mV s$^{-1}$, (c) CV curves at different scan rates from 5 to 50 mV s$^{-1}$, (d) GCD curves at different specific currents from 0.5 to 10.0 A g$^{-1}$.

The Nyquist plots of NiFe-LDHs electrodes (Fig. 6e) showed typical SC characteristic impedance curves [36]. The interception at real axis and the diameter of the semicircle indicate the equivalent series resistance R$_{ESR}$ and the charge transfer resistance R$_{ct}$, respectively [37]. The R$_{ESR}$ and the R$_{ct}$ values are summarized in Table S3 by fitting EIS data with an equivalent circuit with constant phase element (Fig. S6). Compared with bulky Ni$_3$Fe$_1$-LDHs-0, the lower R$_{ESR}$ of Ni$_3$Fe$_1$-LDHs-2 electrode demonstrates the enhanced conductivity. The lowest R$_{ct}$ value of Ni$_3$Fe$_1$-LDHs-2 electrode (0.66 Ω) illustrated that ultrasonic treatment could promote the diffusion of hydroxide ions in the electrolyte and accelerate the kinetics of redox reaction. The cyclability of Ni$_3$Fe$_1$-LDHs-2 electrode at specific current of 10 A g$^{-1}$ after 1000 cycles was supplied in Fig. 6f. The capacitance retention remained 98% without considering the initial activation processes, revealing exceptional cycle stability.

In general, the considerable reasons of the excellent capacitive performance of the LDHs nanosheets could be ascribed to the synergetic effect of appropriate structural parameters, including the ultrathin thickness, large specific surface area, and high mesoporous volume. These structural parameters are controlled by suitable ultrasonic time. Excessive ultrasonic treatment could destroy the structure of LDHs nanosheets, giving rise to deactivation of active sites of charge storage. In this work, 2 h is the most suitable ultrasonic time for NiFe-LDHs nanosheets as electrode materials of SCs. As listed in Table S4, the electrochemical performance of $Ni_3Fe_1$-LDHs-2 is comparable or superior to NiFe-LDHs materials in relative literatures.

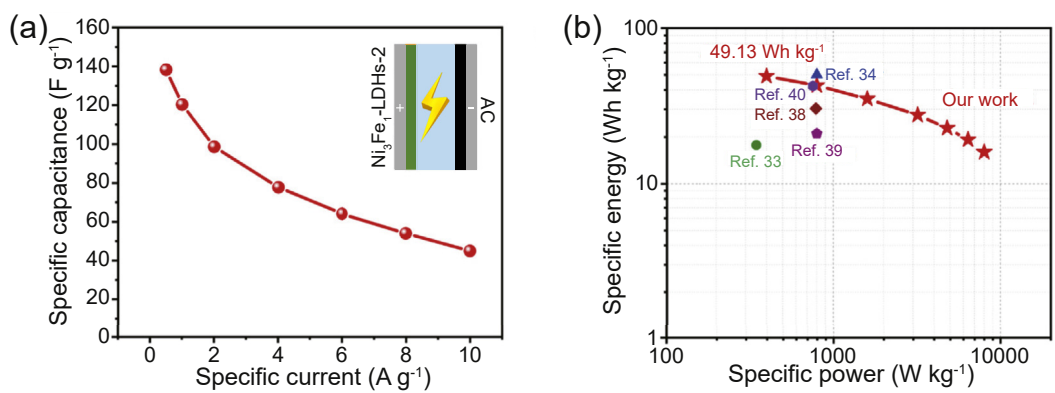

Fig. 8. (a) Specific capacitance of the ASC at different specific currents. (b) Ragone plot related to the specific energy and specific power of the ASC.

### 3.3. Electrochemical characterization of asymmetric supercapacitor

An asymmetric supercapacitor (ASC) device is designed to further demonstrate the potential for practical application of $Ni_3Fe_1$-LDHs electrodes. The ASC device was denoted as $Ni_3Fe_1$-LDHs//AC ASC. Fig. 7a demonstrates the CV curves of the positive electrode ($Ni_3Fe_1$-LDHs-2) and the negative electrode (AC) in the three-electrode configuration. The corresponding potential range was chosen from −0.2 V to 0.5 V and from −1.1 V to −0.2 V to match the voltage window of two-electrode system, which is determined in Fig. 7b. By comparing CV curves of the $Ni_3Fe_1$-LDHs//AC ASC operated in different voltage windows, the stabilized working voltage was chosen as 1.6 V since the obvious polarization phenomenon was observed with the voltage range of 0.0–1.7 V. The capacitive performance of the ASC is evaluated by CV curves at different scan rates of 5–50 mV s$^{-1}$ (Fig. 7c) and GCD curves at different specific currents of 0.5–10.0 A g$^{-1}$ (Fig. 7d). It is well known that the shape of CV curves of the AC electrode is rectangular. The irregular rectangular shape of the CV curves of the ASC revealed the contributions from electrical double layer capacitive behavior and pseudocapacitive behavior. GCD curves with the symmetric shape show the excellent electrochemical reversibility, which originates from the electrical double layer capacitance and pseudocapacitance.

In Fig. 8, the specific capacitance values and Ragone plot of the $Ni_3Fe_1$-LDHs//AC ASC are presented according to GCD curves. The specific capacitance values at 0.5, 1.0, 2.0, 4.0, 6.0, 8.0, and 10.0 A g$^{-1}$ were calculated as 138, 120, 99, 78, 63, 54, 45 F g$^{-1}$ with the capacitance retention of 32.6%. The Ragone plot in Fig. 8b represents the current status of $Ni_3Fe_1$-LDHs//AC ASC for energy storage. The maximum specific energy of 49.13 Wh kg$^{-1}$ was achieved at the specific power of 400 W kg$^{-1}$. In addition, the specific energy of the as-assembled ASC device was retained to 15.98 Wh kg$^{-1}$ at extremely high specific power of 8000 W kg$^{-1}$. Accordingly, the maximum specific energy values are greater or comparable to the recent published literatures about ASC devices, such as NiFe-LDHs/rGO/nickel foam//MC (17.7 Wh kg$^{-1}$ at 348 W kg$^{-1}$) [33], NiFe-LDHs/nickel foam//AC (50.2 Wh kg$^{-1}$ at 800 W kg$^{-1}$) [34], NiAl-LDHs/MnO$_2$//AC (30.4 Wh kg$^{-1}$ at 790 W kg$^{-1}$) [38], NiAl-LDHs/GO//AC (21.0 Wh kg$^{-1}$ at 800 W kg$^{-1}$) [39], NiFe-LDHs/MXene//AC (42.4 Wh kg$^{-1}$ at 758 W kg$^{-1}$) [40].

### 4. Conclusions

In summary, we have successfully synthesized NiFe-LDHs nanosheets with a facile ultrasonication-assisted methodology. For comparison, bulky NiFe-LDHs were also fabricated without ultrasonic treatment. The structure and the composition of NiFe-LDHs were confirmed by XRD and TGA results. The interlayer spacing of NiFe-LDHs was enlarged with the aid of ultrasonic treatment during the synthesis process. The NiFe-LDHs nanosheets prepared with 2-h ultrasonic treatments possessed the ultrathin thickness (7.39 nm), large specific surface area (77.16 m$^2$ g$^{-1}$), and high mesoporous volume (0.080 cm$^3$ g$^{-1}$), leading to the excellent capacitive performance. However, excessive ultrasonication could decrease number of active sites of energy storage. The maximum specific capacitance of $Ni_3Fe_1$-LDHs nanosheets prepared with 2-h ultrasonic treatments reached 1923 F g$^{-1}$, which is almost 1.4-fold and 1.7-fold of that of bulky $Ni_3Fe_1$-LDHs and $Ni_3Fe_1$-LDHs nanosheets prepared with 3-h ultrasonic treatments, respectively. The characterization results clearly demonstrate the complicated relationship between the morphology/thickness and capacitive performance of LDHs with the utilization of ultrasonication. Additionally, the assembled NiFe-LDHs//AC ASC achieved a great specific energy of 49.13 Wh kg$^{-1}$ at 400 W kg$^{-1}$. This work provides a guidance for designing and optimizing the structure of the ultrasonication-assisted exfoliated LDHs nanosheets for SCs.

### Conflict of interest

The authors declare no conflict of interest.

### Acknowledgments

The authors acknowledged the financial supports from Key-Area Research and Development Program of Guangdong Province (2019B110209003), Guangdong Basic and Applied Basic Research Foundation (2019B1515120058, 2020A1515011149), National Key R&D Program of China (2018YFD0800700), National Ten Thousand Talent Plan, National Natural Science Foundation of China (21776324), the Fundamental Research Funds for the Central Universities (19lgzd25), and Hundred Talent Plan (201602) from Sun Yat-sen University.

### Appendix A. Supplementary data

Supplementary data to this article can be found online at https://doi.org/10.1016/j.gee.2021.01.019.

### References

[1] S. Liu, L. Wei, H. Wang, Appl. Energy 278 (2020) 115436.
[2] D. Chen, K. Jiang, T. Huang, G. Shen, Adv. Mater. 32 (2020) 1901806.
[3] Z. Pang, J. Duan, Y. Zhao, Q. Tang, B. He, L. Yu, J. Power Sources 400 (2018) 126–134.
[4] Z. Chen, M. Zhang, Y. Wang, Z. Yang, D. Hu, Y. Tang, K. Yan, Green Energy Environ. 6 (2021) 929–937.

[5] H. Liu, Y. Xie, J. Liu, K.-S. Moon, L. Lu, Z. Lin, W. Yuan, C. Shen, X. Zang, L. Lin, Y. Tang, C.-P. Wong, Chem. Eng. J. 393 (2020) 124672.
[6] J. Wang, N. Wu, T. Liu, S. Cao, J. Yu, Acta Phys. Chim. Sin. 36 (2020) 1907072.
[7] Z. Lu, X. Xu, Y. Chen, X. Wang, L. Sun, K. Zhuo, Green Energy Environ. 5 (2020) 69–75.
[8] S. Yang, S. Wang, X. Liu, L. Li, Carbon 147 (2019) 540–549.
[9] B. Ding, D. Guo, Y. Wang, X. Wu, Z. Fan, J. Power Sources 398 (2018) 113–119.
[10] H. Liu, X. Liu, S. Wang, H.-K. Liu, L. Li, Energy Storage Mater. 28 (2020) 122–145.
[11] M. Inagaki, H. Konno, O. Tanaike, J. Power Sources 195 (2010) 7880–7903.
[12] R. Patel, J.T. Park, M. Patel, J.K. Dash, E.B. Gowd, R. Karpoormath, A. Mishra, J. Kwak, J.H. Kim, J. Mater. Chem. A 6 (2018) 12–29.
[13] Q. Yang, Z. Li, R. Zhang, L. Zhou, M. Shao, M. Wei, Nano Energy 41 (2017) 408–416.
[14] Y. Ouyang, T. Xing, Y. Chen, L. Zheng, J. Peng, C. Wu, B. Chang, Z. Luo, X. Wang, J. Energy Storage 30 (2020) 101454.
[15] X. Li, D. Du, Y. Zhang, W. Xing, Q. Xue, Z. Yan, J. Mater. Chem. A 5 (2017) 15460–15485.
[16] T. Li, X. Hao, S. Bai, Y. Zhao, Y.-F. Song, Acta Phys. Chim. Sin. 36 (2020) 1912005.
[17] R. Li, Y. Liu, H. Li, M. Zhang, Y. Lu, L. Zhang, J. Xiao, F. Boehm, K. Yan, Small Methods 3 (2019) 1800344.
[18] J. Cui, Z. Li, G. Wang, J. Guo, M. Shao, J. Mater. Chem. A 8 (2020) 23738–23755.
[19] X. Gao, P. Wang, Z. Pan, J.P. Claverie, J. Wang, ChemSusChem 13 (2020) 1226–1254.
[20] H. Yin, Z. Tang, Chem. Soc. Rev. 45 (2016) 4873–4891.
[21] Z. Yi, C. Ye, M. Zhang, Y. Lu, Y. Liu, L. Zhang, K. Yan, Appl. Surf. Sci. 480 (2019) 256–261.
[22] Y. Liu, M. Zhang, D. Hu, R. Li, K. Hu, K. Yan, ACS Appl. Energy Mater. 2 (2019) 1162–1168.
[23] Y. Zhao, Q. Wang, T. Bian, H. Yu, H. Fan, C. Zhou, L.-Z. Wu, C.-H. Tung, D. O'hare, T. Zhang, Nanoscale 7 (2015) 7168–7173.
[24] Z. Yang, X. Wang, H. Zhang, S. Yan, C. Zhang, S. Liu, ChemElectroChem 6 (2019) 4456–4463.
[25] N. Mao, C.H. Zhou, D.S. Tong, W.H. Yu, C.X. Cynthia Lin, Appl. Clay Sci. 144 (2017) 60–78.
[26] Z. Liu, R. Ma, M. Osada, N. Iyi, Y. Ebina, K. Takada, T. Sasaki, J. Am. Chem. Soc. 128 (2006) 4872–4880.
[27] T.S. Munonde, H. Zheng, P.N. Nomngongo, Ultrason. Sonochem. 59 (2019) 104716.
[28] Z. Chen, H. Deng, M. Zhang, Z. Yang, D. Hu, Y. Wang, K. Yan, Nanoscale Adv. 2 (2020) 2099–2105.
[29] M. Zhang, Y. Liu, B. Liu, Z. Chen, H. Xu, K. Yan, ACS Catal. 10 (2020) 5179–5189.
[30] B. Liu, M. Zhang, Y. Wang, Z. Chen, K. Yan, J. Alloys Compd. 852 (2021) 156949.
[31] M. Li, F. Liu, X.B. Zhang, J.P. Cheng, Phys. Chem. Chem. Phys. 18 (2016) 30068–30078.
[32] Y. Wang, H. Dou, J. Wang, B. Ding, Y. Xu, Z. Chang, X. Hao, J. Power Sources 327 (2016) 221–228.
[33] M. Li, R. Jijie, A. Barras, P. Roussel, S. Szunerits, R. Boukherroub, Electrochim. Acta 302 (2019) 1–9.
[34] Y. Lu, B. Jiang, L. Fang, F. Ling, J. Wu, B. Hu, F. Meng, K. Niu, F. Lin, H. Zheng, J. Alloys Compd. 714 (2017) 63–70.
[35] R.R. Salunkhe, Y.V. Kaneti, Y. Yamauchi, ACS Nano 11 (2017) 5293–5308.
[36] Y. Wang, X. Qiao, C. Zhang, X. Zhou, J. Energy Storage 26 (2019) 100968.
[37] X. Deng, Y. Jiang, Z. Wei, M. Mao, R. Pothu, H. Wang, C. Wang, J. Liu, J. Ma, Green Energy Environ. 4 (2019) 382–390.
[38] W. Zheng, S. Sun, Y. Xu, R. Yu, H. Li, J. Alloys Compd. 768 (2018) 240–248.
[39] L. Zhang, H. Yao, Z. Li, P. Sun, F. Liu, C. Dong, J. Wang, Z. Li, M. Wu, C. Zhang, B. Zhao, J. Alloys Compd. 711 (2017) 31–41.
[40] H. Zhou, F. Wu, L. Fang, J. Hu, H. Luo, T. Guan, B. Hu, M. Zhou, Int. J. Hydrogen Energy 45 (2020) 13080–13089.